\documentclass[a4paper]{jpconf}
\usepackage{graphicx}
\usepackage{amssymb,amsmath}
\usepackage{bm}
\usepackage{subfigure}

\begin{document}
\title{Particle MCMC for Bayesian Microwave Control}

\author{ P. Minvielle$^1$, A. Todeschini$^2$, F. Caron$^3$, P. Del Moral$^{4,}$}
\address{$^1$ CEA-CESTA, 33114 Le Barp, France}
\address{$^2$ INRIA Bordeaux Sud-Ouest, 351, cours de la Liberation, 33405 Talence Cedex, France}
\address{$^3$ University of Oxford, 1 South Parks Road, Oxford, UK}
\address{$^4$ UNSW, High Street, Kensington Sidney, Autralia}
\ead{pierre.minvielle@cea.fr}

\begin{abstract}
We consider the problem of local radioelectric property estimation from global electromagnetic scattering measurements. This challenging ill-posed high dimensional inverse problem can be explored  by intensive computations of a parallel Maxwell solver on a petaflopic supercomputer.  Then, it is shown how  Bayesian inference can be perfomed with a Particle Marginal Metropolis-Hastings (PMMH) approach, which includes a Rao-Blackwellised Sequential Monte Carlo algorithm with interacting Kalman filters. Material properties, including a multiple components "Debye relaxation"/"Lorenzian resonant" material model, are estimated; it is illustrated on synthetic data. Eventually, we propose different ways to deal with higher dimensional problems, from parallelization to the original introduction of efficient sequential data assimilation techniques, widely used in weather forecasting, oceanography, geophysics, etc.
\end{abstract}

\section{Introduction}
Unlike usual electromagnetic (EM) material characterization techniques \cite{knott2004radar}, the microwave control problem involves to determine or check radioelectric properties (i.e. relative dielectric permittivity and magnetic permeability) of materials that are assembled and placed on the full-scaled object or system, from global scattering measurements (Radar Cross Section) \cite{giraud2013advanced}.

An axisymmetrical object or mock-up is illuminated by a monostatic radar that fulfills to a certain extent directivity and far-field conditions \cite{minvielle2011sparse}. It illuminates the object at a given incidence with a quasi-planar monochromatic continuous wave (CW) of frequency $f$, the object backscatters a CW to the radar  at the same frequency. With an appropriate  instrumentation system (radar, network analyzers, etc.) and a calibration process, it is possible to measure the complex scattering coefficient. It sums up the EM scattering, indicating the wave change in amplitude and phase. It quantifies a global characteristic of the whole object-EM wave interaction in specific conditions (incidence, frequency, etc.). The scattering coefficient are measured for different transmitted and received polarizations. Eventually, various complex scattering coefficients $\mathcal{S}$ are measured at different wave frequencies ($f\in\{f_1,f_2,\cdots,f_{K_f}\}$, for $K_f$ successive discrete frequencies) from a SFCW (Stepped Frequency Continuous Wave) burst,  at different incidence angles ($\theta \in\{\theta_1,\theta_2,\cdots,\theta_{K_{\theta}}\}$, for $K_\theta$ different incidence angles) where the object is rotated with a motorized rotating support, at different transmitted/received linear polarizations ($\mathrm{pol}\in\{HH,VV\}$) \footnote{Notice that HV and VH cross-polarization scattering coefficients can not be considered since they are null (due to the object axisymmetry and the trajectory of the electric and magnetic fields).}. Otherwise, let assume that the object is  axisymmetric and made of  one metallic material, with its associated isotropic radioelectric properties weakly varying .The aim is to determine, from the global scattering measurement  $\mathcal{M}$, the unknown isotropic local EM properties $(\epsilon_1,\mu_1), (\epsilon_2,\mu_2), \cdots, (\epsilon_N,\mu_N)$ along the object,  where $N$ is the number of different elementary areas.

\begin{figure}[ht!] \centering 
\subfigure{\includegraphics*[width=0.54\textwidth,clip=true,viewport=0cm 6cm 28cm 15cm]{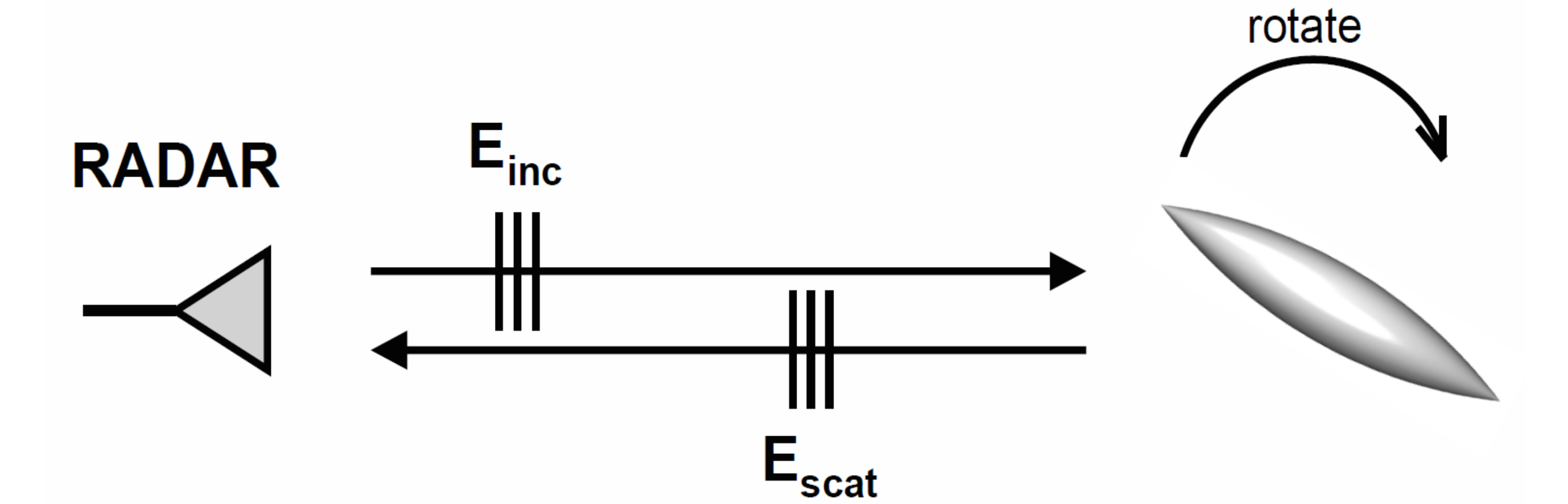}}
\subfigure{\includegraphics*[width=0.45\textwidth,clip=true,viewport=6cm 7cm 22cm 14cm,draft=false] {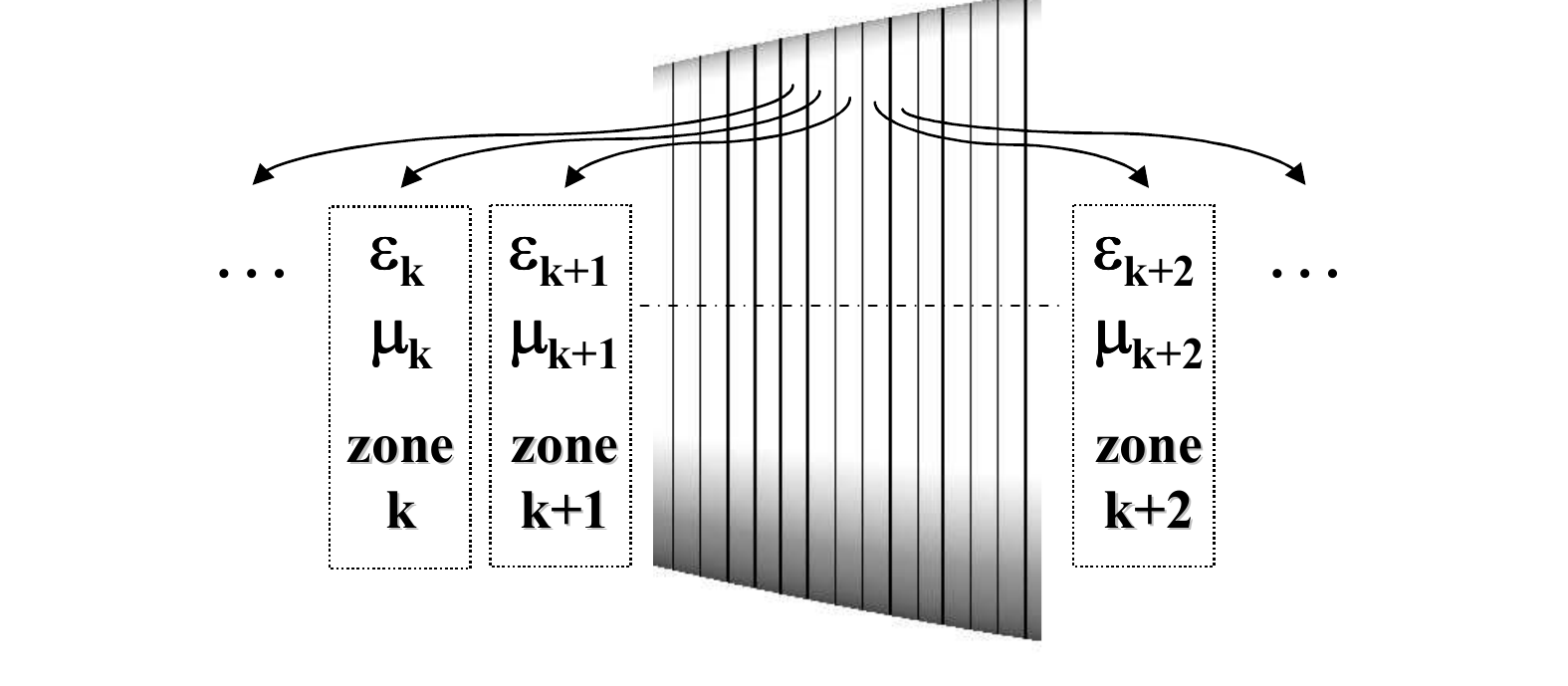}}
\caption{RCS measurement setup (left) -  Elementary mesh zones (right)} \label{Shape} 
\end{figure}

\begin{figure}[ht!] \centering{\includegraphics[width=0.8\linewidth,clip=true,viewport=0cm 7cm 29cm 14cm]{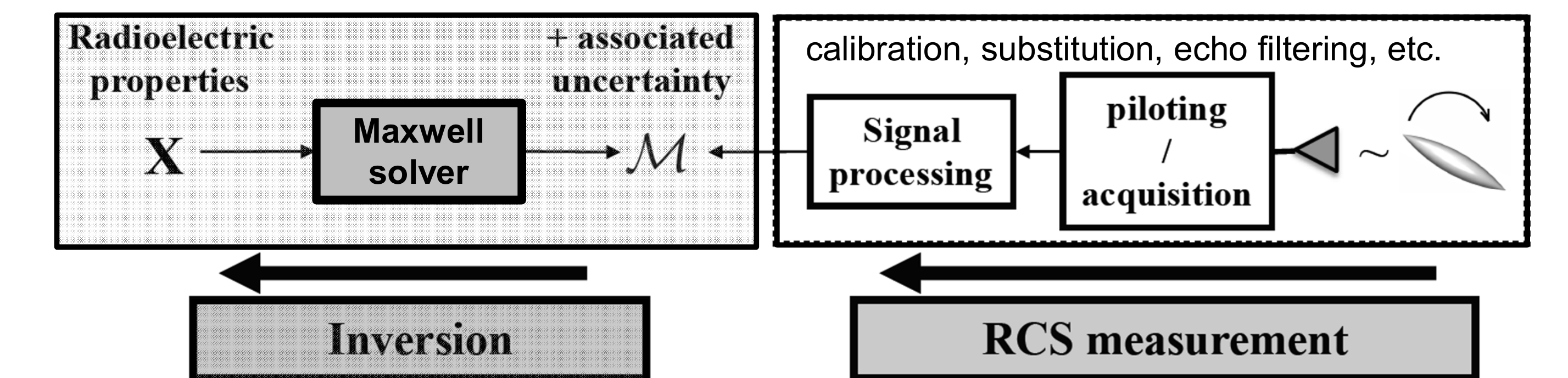} \caption{High dimensional  inverse problem: "row upstream" the Maxwell solver} \label{InvPb}}\end{figure}

It is shown in \cite{giraud2013advanced} that it can lead to a high dimensional  inverse problem, that requires to go upstream a parallelized harmonic Maxwell solver (volume finite element/integral equation) \cite{stupfel1991combined}. Figure \ref{InvPb} sums up the entire inverse scattering problem. On the one hand, the RCS measurement process, that includes acquisition, signal processing, calibration, etc., provides the complex scattering measurement $\mathcal{M}$, with uncertainties. On the other hand,  it would be useful to "row upstream" the Maxwell solver, in order to determine the unknown radioelectric properties, denoted by $\mathbf{x}$. Yet, even with recourse to HPC, there is no direct way to solve what turns out to be a high dimensional ill-posed inverse problem. On the contrary, the forward scattering model based on the resolution of Maxwell's equations can determine the scattering coefficients, given the EM properties, the object geometry and acquisition conditions (i.e. wave frequency, incidence, etc.). It  lies in the resolution of Maxwell's equations, partial derivative equations that represent the electromagnetic scattering problem of an inhomogeneous obstacle.  It is performed by an efficient parallelized harmonic Maxwell  solver, an exact method that combines a volume finite element method and integral equation technique,  taking benefit from the axisymmetrical geometry of the shape \cite{stupfel1991combined}. Discretization is known to lead to problems of very large sizes, especially when the frequency is high. 

This current work is an extension to \cite{giraud2013advanced}. A Bayesian inference approach, based on "particle MCMC", is developed.  It can perform estimation of material properties and determine a multiple components (Debye relaxation/Lorenzian resonant) material model. It provides various ways to deal with higher dimensional problems, from massively parallel computing to high-dimension oriented adaptations.

\section{Problem statement : inference on a general HMM}

\begin{figure}[ht!] \centering{\includegraphics[width=0.6\linewidth]{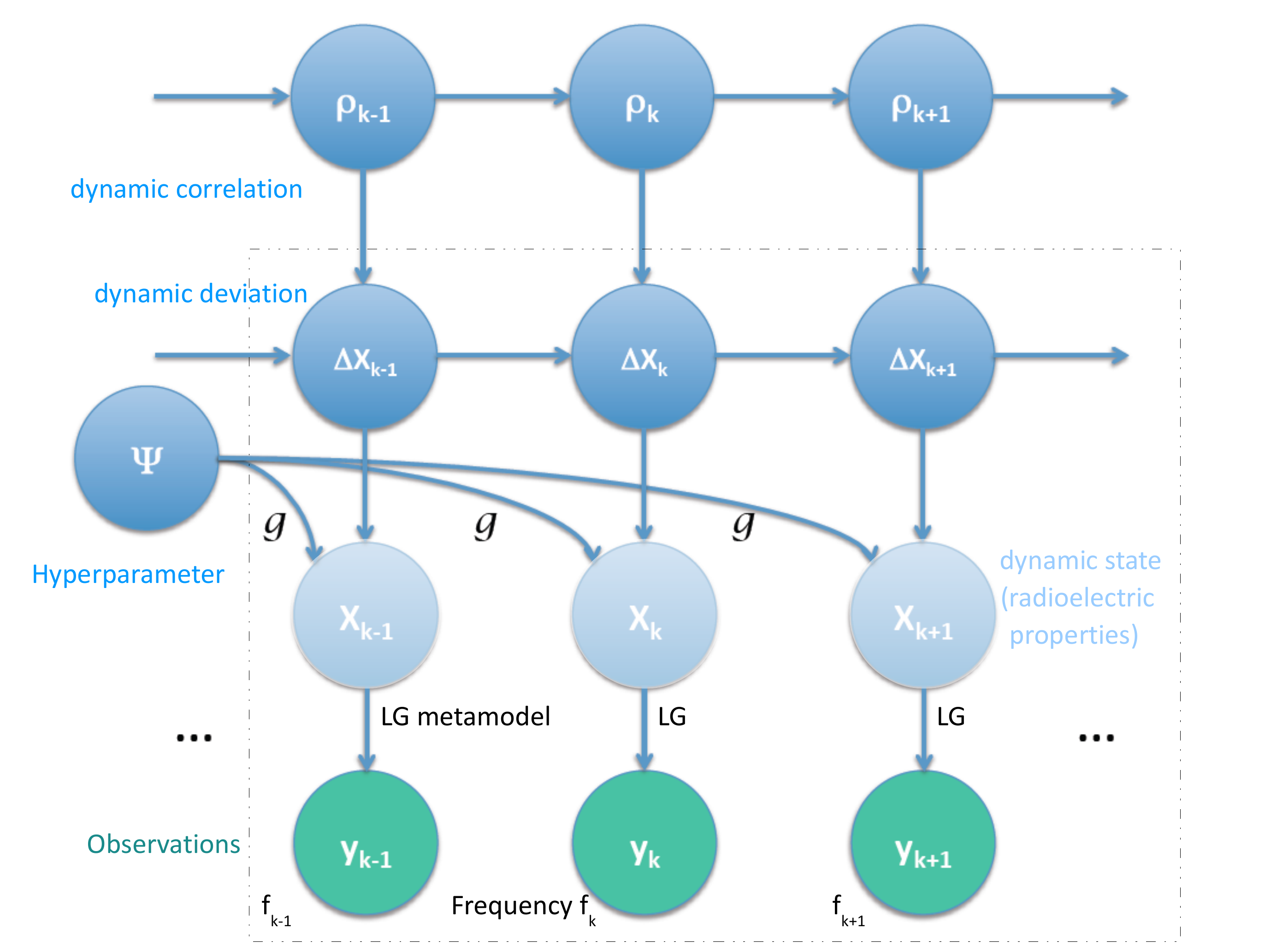}\caption{General Hidden Markov Model} \label{ModG_Ext2}}\end{figure}

The problem statement can be described as a general HMM (see figure \ref{ModG_Ext2}). The graphical model is globally composed of a hierarchy of hidden states, at successive frequencies $f_k$, of a fixed hyperparameter $\bm{\Psi}$ (in blue) and of observations. Next, we detail the various items. 

\paragraph{State: } The state $\mathbf{x_k}=\left[  	\bm{\underline{\epsilon}_k}^{\prime}  \quad \bm{\underline{\epsilon}_k}^{\prime\prime}  \quad \bm{\underline{\mu}_k}^{\prime}  \quad \bm{\underline{\mu}_k}^{\prime\prime} \right]^T$ is composed of the real ($\prime$) and imaginary  ($\prime\prime$) radioelectric components at frequency $f_k$ ($N$ elementary areas).  It includes all the unknown EM properties that are to be estimated (at frequency $f_k$), i.e. the relative permittivity and permeability components of the $N$ elementary zones  (at frequency $f_k$). Omitting the indice $k$, the  four components can be developed  as: $\bm{\underline{\epsilon}}^{\prime}=\left[\epsilon^{\prime}_1  \cdots \epsilon^{\prime}_N \right]^T$, $
\bm{\underline{\epsilon}}^{\prime\prime}=\left[\epsilon^{\prime\prime}_1  \cdots \epsilon^{\prime\prime}_N \right]^T$, $
\bm{\underline{\mu}}^{\prime}=\left[\mu^{\prime}_1  \cdots \mu^{\prime}_N \right]^T$ and $
\bm{\underline{\mu}}^{\prime\prime}=\left[\mu^{\prime\prime}_1  \cdots \mu^{\prime\prime}_N \right]^T$. Consequently, the state  $\mathbf{x_k}$ is in a system space of  dimension $4N$. Here, we are specially interested in the following decomposition: 
\begin{equation}
	 \mathbf{x}_k=g(f_k,\bm{\Psi})+\Delta\mathbf{x}_k
\end{equation}
In this model, the EM properties  $\mathbf{x_k}$ are supposed to partly follow a deterministic physical material model $g$ and partly a random deviation term  $\Delta\mathbf{x}_k$. The parametric material model $g(f_k,\bm{\Psi})$ depends on the frequency, where  $\bm{\Psi}$ is the associated unknown hyperparameter. It is a sum of Debye relaxation/Lorenzian resonant terms (see \cite{baker1992nonlinear} for details).The deviation (from model) $\Delta\mathbf{x}_k$ can be modeled as an AutoRegressive AR(1) model given frequential correlation $\rho_k$ (modelled by a Markov process related to a random walk): $\Delta\mathbf{ x}_1  \sim  \mathcal{N} \left( \mathbf{0}, \mathbf{P}_1\right)$ and $\Delta\mathbf{ x}_{k+1}=\mathbf{M}_k^{\rho}\cdot\Delta\mathbf{ x}_k+\mathbf{w}_k$, where $\mathbf{w}_k$: Gaussian  noise ($\mathbb{E}(\mathbf{w}_k)\neq0$). Notice that this stochastic process  includes also spatial correlation (see \cite{giraud2013advanced} for details). 

\paragraph{Observation: } $\mathbf{y_k}=[\cdots]^T$ is composed of the complex scattering coefficients,  measured at frequency $f_k$, for  various rotation angles $\theta_1, \cdots, \theta_{K_\theta}$. 

\paragraph{Likelihood model: } The following likelihood model can be learned from intensive Maxwell solver computations on a petaflopic supercomputer at each frequency $f_k$. The high-dimensional space is explored by random sampling, according to a prior knowledge about
the expected EM properties. Then, multidimensional linear regression and RCS measurement modeling (Gaussian additive noise with covariance matrix $\mathbf{R_k}$) lead to the following Linear Gaussian metamodel  (valid in a limited domain of interest \cite{giraud2013advanced}, fitness can be accessed by residual analysis):
\begin{equation}
\mathbf{y}_k=\left[\mathbf{A}_k \cdot \mathbf{x}_k + \mathbf{y}_k^0 \right]+
\mathbf{v}_k \textrm{, } \mathbf{v}_k \sim \mathcal{N} (0,\mathbf{R_k})
\end{equation}

\paragraph*{} The problem is to estimate jointly the fixed hyperparameter $\bm{\Psi}$, the dynamic states $\Delta\mathbf{x}_1, \cdots, \Delta\mathbf{x}_K$ and the dynamic frequency correlations $\rho_1, \cdots, \rho_K$, from the measurements $\mathbf{y_1},\mathbf{y_2},\cdots,\mathbf{y_K}$ (noticing that the general HMM contains a specific structure, i.e. the conditionally Linear Gaussian property given $\bm{\Psi}$ and $\rho_1,\cdots,\rho_K$). Bayesian estimation provides more information than standard maximum likelihood techniques (gradient or Expectation Maximization based).

\section{Particle MCMC : PMMH (with Interacting KF)}
\subsection{Principle}
Bayesian inference, in the general HMM context, can be performed by  recent and powerful approaches, called ``Particle MCMC'' \cite{andrieu2010particle}. Since they involve two types of samplers, known as MCMC (Markov Chain Monte Carlo) and SMC (Sequential Monte Carlo), they are known to be computationally expensive. Here, we consider the specific method "Particle Marginal Metropolis-Hastings" (PMMH). In our context, it can solve the "smoothing" and parameter estimation simultaneously,  managing to sample from the joint posterior distribution $p(\bm{\Psi},\Delta\mathbf{x}_{1:K}, \rho_{1:K} | \mathbf{y_{1:K}})$. Is is based on a designed  PMMH Markov chain which invariant distribution is the target joint distribution.

\begin{figure}[ht!] \centering{\includegraphics[angle=270,width=0.9\linewidth,clip=true,draft=false,viewport=3cm 3.cm 18cm 27.5cm]{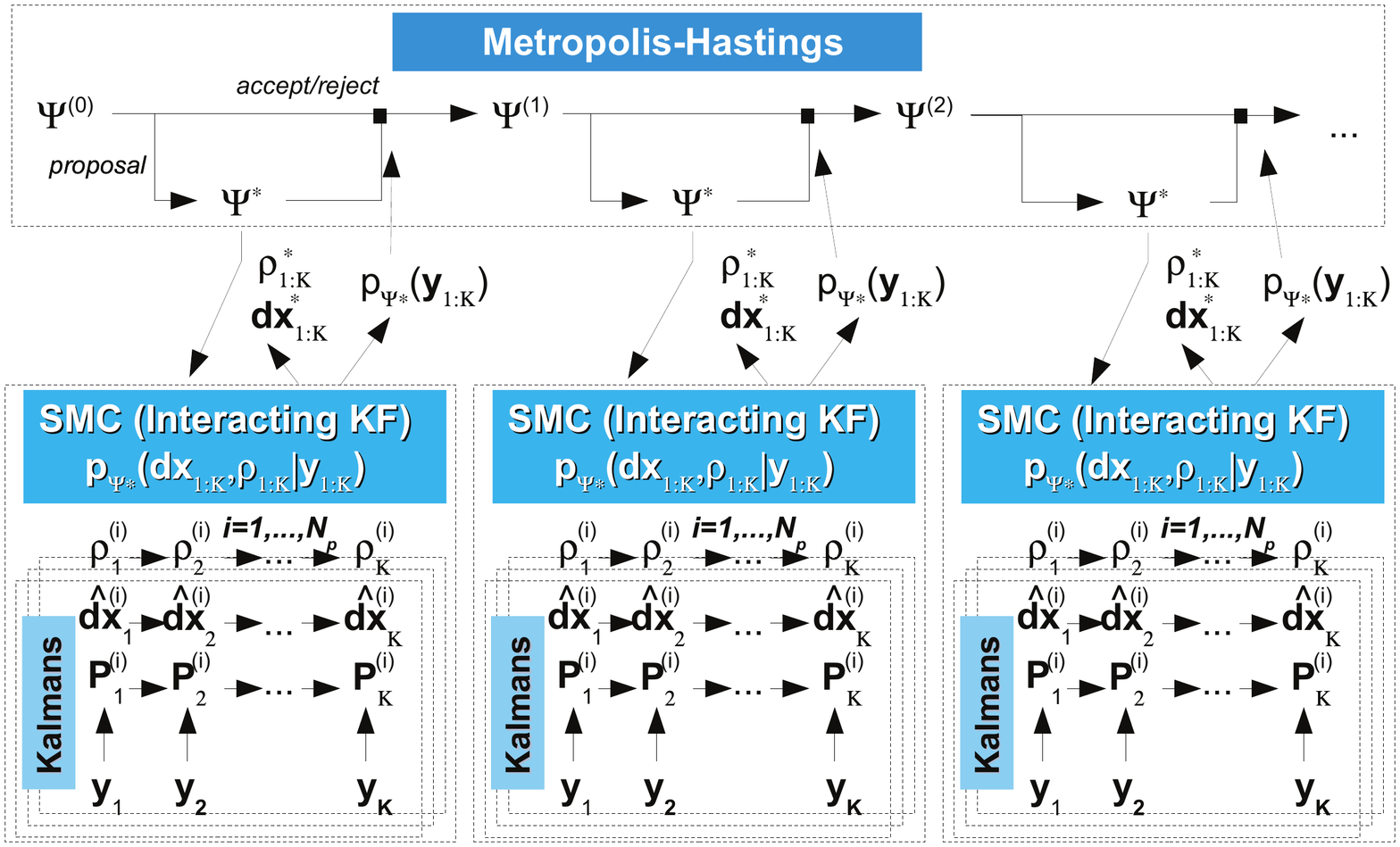}\caption{ Rao-Blackwellised PMMH (with Interacting KF) scheme}  \label{Schem1}} \end{figure}

Close to \cite{nevat2010channel}, our approach for the microwave control application lies on a PMMH multilevel stochastic algorithm. It is formed of 2 embedded levels (see figure \ref{Schem1}):
\begin{itemize}
\item \textbf{MCMC higher level [Metropolis-Hastings]:} the Metropolis-Hastings methods is able to sample from $\bm{\Psi}^\star$.
\item \textbf{SMC lower level [Rao-Blackwellized]: } A variance reduction strategy leads to a Rao-Blackwellised SMC method. It consists in a bank of interacting Kalman filters\cite{doucet2000sequential}, that are able to compute the marginal likelihood  $p(\mathbf{y_{1:K}}|\bm{\Psi}^\star)$ and sample from $(\Delta\mathbf{x}_{1:K}^\star,\rho_{1:K}^\star)$.
\end{itemize}

\subsection{Illustration (synthetic data)}
\begin{figure}[ht!] \centering{\includegraphics[angle=0,width=1\linewidth,clip=true,viewport=2cm 2cm 27.5cm 19cm,draft=false]{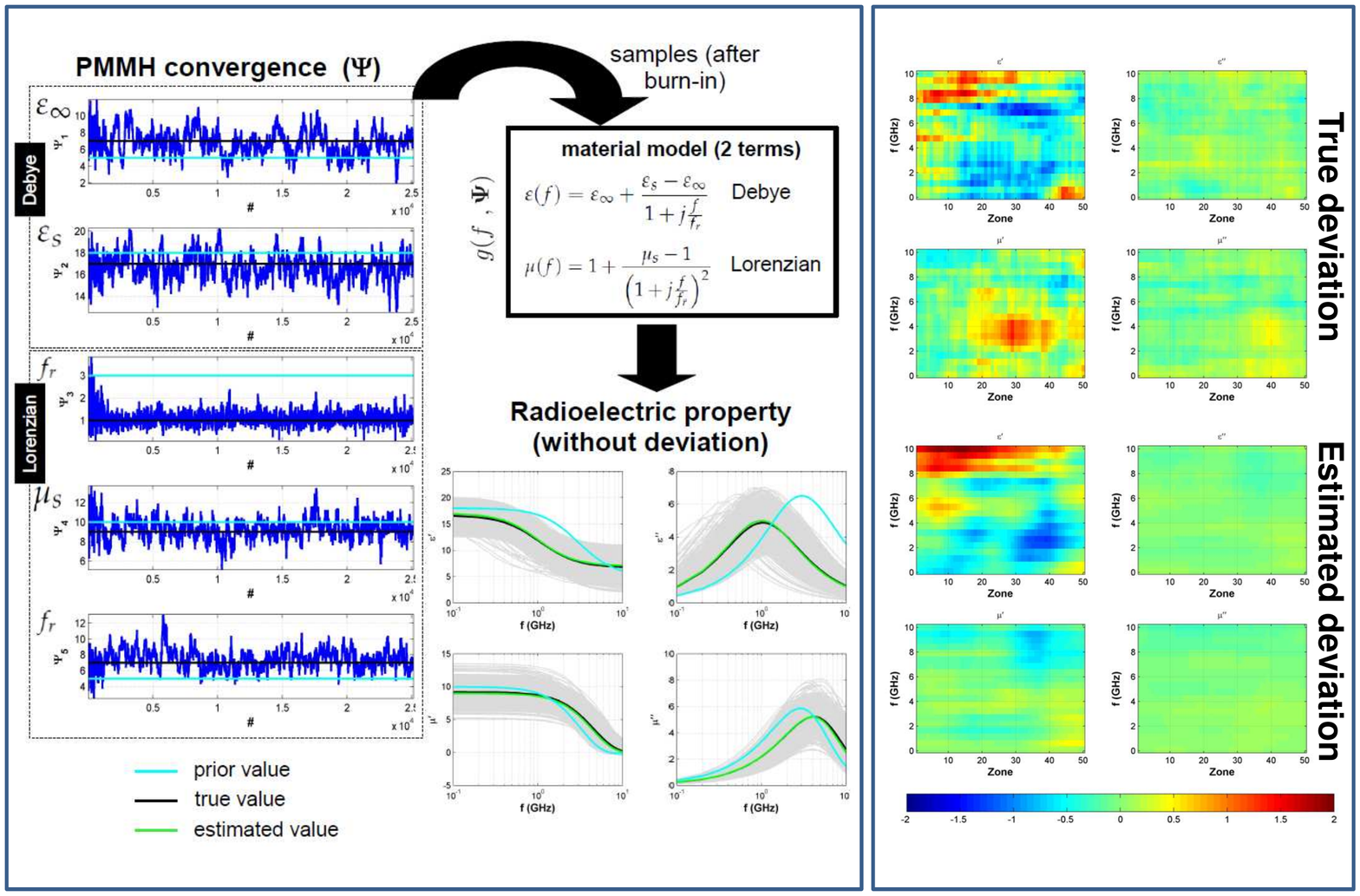} \caption{Results: chain convergence (left) - Deviations (right)}  \label{RES}} \end{figure}

\paragraph{Assumptions: } The PMMH approach is applied on synthetic data, the dimension of which is relatively low. The object is composed of one material (dim($\mathbf{x_k}$)=$50\times4$), with spatial inhomogeneity. The associated material model (in frequency) is made of 2 terms (mentioned in figure \ref{RES}): a Debye term (parameters: $\epsilon_\infty$ and $\epsilon_s$ ) and Lorenzian term (parameters: $f_r$ , $\mu_s$ and $f_r$). The deviation is simulated from the AR process model. Concerning the measurements, we consider $K=20$ ($f\in$ [0.1-10 GHz]), $K_\theta=100$ ($\theta\in [0^\circ-180^\circ]$), dim($\mathbf{y_k}$)=$100\times4$.

\paragraph{Inference process: }For efficiency concerns, an adaptive PMMH algorithm  \cite{peters2010ecological} was developed. The MCMC level includes a tempering phase and a kernel mixture strategy. The SMC level is based on the classical SIR (Sampling Importance Resampling) algorithm, with one hundred ``Kalman'' particles. The whole PMMH behavior is illustrated in the left part figure \ref{RES}. After a burn-in phase, the algorithm provides samples that approximate the posterior distribution. Propagated through the material model, the samples can be used to predict the deterministic part of the radioelectric properties. They are coherent with the true one. The right part of figure \ref{RES} compares the true deviation and the estimated one.It shows that the important deviations in $\mu^\prime$ are detected while the insignificant ones can not be estimated.

\subsection{Practical issues}
The first practical issue turns out to be the high variation of marginal likelihood noisy estimates, due to a quite high dimension HMM (state dimension). It is solved by the adaptive strategy \cite{peters2010ecological} and by tuning the SMC particle number (control the output noise on $p(\mathbf{y_{1:K}}|\bm{\Psi}^\star)$ and $(\Delta\mathbf{x}_{1:K}^\star,\rho_{1:K}^\star)$), versus the MH step number, the proposal (control of the acceptance rate), etc. Let mention that another way could be to apply another related multilevel stochastic method, called SMC$^2$ \cite{chopin2013smc2}. And yet, the main practical issue is the expected computationally time expensiveness, $\sim$ 1 week on a standard PC despite the metamodeling speeding up. It will increase for higher state and observation dimensions. One way to overcome this issue is to turn towards massively parallel computing (at the SMC level), an important trend in Bayesian computational statistics. Another way hereinafter developed is to introduce high-dimension oriented adaptations and faster approximations.  

\section{High dimension adaptation: PMMH (with Interacting EnKF)} 
To accelerate PMMH inference, we have turned towards Ensemble Kalman Filter (EnKF) \cite{evensen2003ensemble}. It is a Monte-Carlo alternative to (Extended) Kalman filter for huge dimensional state vector. It is widely used in sequential data assimilation: weather forecasting, oceanography, reservoir modelling, etc. It leads to a new original approach: a PMMH algorithm with Rao-Blackwellised SMC based on Interacting EnKF. The principle is to substitute Kalman filters by Ensemble Kalman filters with an efficient implementation, i.e. no empirical covariance manipulation and adaptation to a large number of data points (Sherman-Morrison-Woodbury formula). It provides similar results on dimension-limited problems, such as the above illustration. Concerning the complexity, the benefit is important for large state and observation dimensions (i.e. much higher than the herein problem). Notice that it is also compatible with massively parallel computing.
The theoretical concern is that there is  no longer unbiased estimate of unnormalized target density, required by  PMMH to be an "exact approximation" of idealized MCMC (see \cite{andrieu2010particle}). Yet, in a practical viewpoint, the unbiased condition is actually unattainable, due to various bias sources, from RCS measurements (residual interfering echoes) to the approximate likelihood model.

\section{Conclusion}
A global Bayesian inference approach, consisting in a Particle Marginal Metropolis-Hastings (PMMH) that includes interacting Kalman filters, is developed for micro-wave material control. Based on two sampling levels, both MCMC and SMC, it  simultaneously estimates  model parameters and spatial/frequency deviations, and the associated uncertainties. It is at the cost of a high computational time that increases with dimensions, but can be reduced with massively parallel computing or/and inference approximations from sequential data assimilation.

\section*{References}
\bibliographystyle{iopart-num}
\bibliography{ref}

\end{document}